\documentstyle[12pt]{article}
\begin{document}
\thispagestyle{empty}

\begin{center}
\LARGE \tt \bf{Cosmological backgroung torsion limits from Lorentz violation}
\end{center}

\vspace{2.5cm}

\begin{center} {\large L.C. Garcia de Andrade \footnote{Departamento de F\'{\i}sica Teorica - UERJ

Rua S\~{a}o Fco. Xavier 524, Rio de Janeiro, RJ

Maracan\~{a}, CEP:20550-003 , Brazil.

E-Mail.: garcia@dft.if.uerj.br}}
\end{center}

\vspace{2.0cm}

\begin{abstract}
Cosmological limits on Lorentz invariance breaking in Chern-Simons $(3+1)-dimensional$ electrodynamics are used to place limits on torsion. Birefrigence phenomena is discussed by using extending the propagation equation to Riemann-Cartan spacetimes instead of treating it in purely Riemannian spaces. The parameter of Lorentz violation is shown to be proportional to the axial torsion vector which allows us to place a limit on cosmological background torsion from the Lorentz violation constraint which is given by $ 10^{-33} eV <|S^{\mu}| < 10^{-32} eV$  where $|S^{\mu}|$ is the axial torsion vector.
\end{abstract}

\newpage

Recently some interesting papers have appeared in the literature concerning the determination of torsion limits by making using of a Chern-Simons (CS) type electrodynamics in spacetimes with torsion \cite{1,2,3}. Previous investigation of these electrodynamics include the determination of Lorentz violation limits by Carroll,Field and Jackiw \cite{4} and the more recent investigation of the Birefrigence phenomenon proposed by Nodland and Ralston (NR) \cite{5}. Indeed Carroll et al \cite{6} have shown along with other authors \cite{7,8} that Birefrigence result has not yet been discovered due to a misleading statistical procedure by NR. Earlier Prasana \cite{9} have shown that there were no Birefrigence in Einstein's gravity since photons follow null geodesics irrespective of their polarisation. In Dobado and Maroto \cite{3} paper they have placed limits on torsion by making use of a CS theory and birefrigence proposing that torsion propagating theories of gravity could be used to explain birefrigence itself. More recently Garcia de Andrade \cite{10} have shown similar results without making use of radiative corrections. The limits obtained by Dobado,Maroto and Garcia de Andrade are respectively $10^{-32} eV$ and $10^{-29} eV$. These torsion limits are well within another limit computed by Mohanty and Sarkar \cite{2} on the basis of constraining background torsion fields from the $K_{L}$ and $K_{S}$ mass difference, CPT violating mass difference and CP violating  quantities ${\epsilon}$ and ${\eta}_{+-}$. Their most stringent limit on the cosmological background torsion $<T^{0}> < 10^{-25} GeV$ which comes directly from the measurement of the CPT violation. The Lagrangean of CS eletrodynamics where the minimal coupling with torsion is taken and to simplify matters since we are interested mainly on the torsion effects we not consider Riemannian curvature effects. I also discuss the physical interpretation of the relation of the Lorentz limit and torsion and some discussion on birefrigence hypothesis is undertaken.

Let us now consider the CS electrodynamics action expression

\begin{equation}
S_{CS} = \int{dx^{4}[-(\frac{1}{4}F^{2}+\frac{1}{2}{F^{*}}^{{\mu}{\nu}}A_{{\nu}}p_{\mu})]}
\label{1}
\end{equation}
where ${F^{*}}^{{\mu}{\nu}}={\epsilon}^{{\mu}{\nu}{\alpha}{\beta}}F_{{\alpha}{\beta}}$  is the dual of the electromagnetic field $F_{{\alpha}{\beta}}={\partial}_{\alpha}A_{\beta}-{\partial}_{\beta}A_{\alpha}$  the second term on the RHS of equation (\ref{1}) is the CS term and $p_{\alpha}$ is the mass parameter that respects gauge invariance while it is the Lorentz invariance which is violated. Instead of starting from this action we considered the Proca electrodynamics action without torsion in Minkowski space 

\begin{equation}
S_{P} = \int{dx^{4}[-\frac{1}{4}F^{2}+\frac{{m_{\gamma}}^{2}}{2}A^{\nu}A_{\nu}]}
\label{2}
\end{equation}
where $ A^{\nu}$ is the electromagnetic  vector potential and  ${m_{\gamma}}^{2}$ is the squared mass of the photon.Let us now consider the minimal coupling perscription ${\partial}_{\mu}A_{\nu}\rightarrow {\partial}_{\mu}A_{\nu}+ {S_{{\mu}{\nu}}}^{\lambda}A_{\lambda}$. Substituting this expression into the Maxwell electromagnetic field tensor $F_{{\mu}{\nu}}$ this transformation into expression (\ref{2}) allows us to obtain the following expression
\begin{equation}
\frac{1}{4}{F'}^{2}=-\frac{1}{4}F^{2}-\frac{1}{2}{S_{{\nu}{\lambda}}}^{\alpha} A_{\alpha}F^{{\nu}{\lambda}}-{S_{{\nu}{\lambda}}}^{\beta} {S^{{\nu}{\lambda}}}_{\alpha} A_{\beta}A^{\alpha}
\label{3}
\end{equation}
Substitution of this result into the expression of the Proca action one notices that the last term on the RHS of the expression (\ref{3}) maybe cancelled allowing the photon to have a mass generated by torsion squared. Besides we are left with an action of the NR type which implies that the mass parameter of Lorentz violation becomes

\begin{equation}
-\frac{1}{2}{p_{\alpha}}= S_{\alpha}
\label{4}
\end{equation}
where this result has been obtained by making use of the axial torsion vector which as we shall see later leads to an axionic form of torsion depending of a scalar field. This is in fact as shown before an explanation for the fact that the cosmic rotation axis is written in terms of the gradient of a scalar field. To resume this first section we may say that we are left with a massive photon where the explicitly putted by hand Proca massive term dissappears. Let us now consider explicitly the field equations obtained from the action we were left with. The CS form of the field equations is simply obtained by the substitution of the current $J^{\alpha} \rightarrow J^{\alpha}+ p_{\mu} {F^{*}}^{{\mu}{\alpha}}$ and therefore by expressing the field equations in terms of the torsion axial vector $S^{\alpha}$ one obtains 
\begin{equation}
{\partial}_{\mu}F^{{\mu}{\nu}}=4{\pi}J^{\nu}- 2S_{\mu} {F^{*}}^{{\mu}{\nu}}
\label{5}
\end{equation}
where in terms of components 
\begin{equation}
div \vec{E}=4{\pi}{\rho}+ 2\vec{S}\vec{B}
\label{6}
\end{equation}
and 

\begin{equation}
-{\partial}_{t}\vec{E} + {\nabla}X \vec{B}= 4{\pi}\vec{J}+2S_{0}\vec{B}-2\vec{S}X \vec{E}
\label{7}
\end{equation}
The homogeneous Maxwell equations are obtained by simply applying
\begin{equation}
div \vec{B}=0
\label{8}
\end{equation}
and
\begin{equation}
{\partial}_{t}\vec{B} + {\nabla} X \vec{E}=0
\label{9}
\end{equation}

Of course one must note that in the absence of torsion vector we reduce our equations to Maxwell equations in flat spacetime. Earlier \cite{10} I have discussed in detailed the electrodynamics with massive photons with torsion and found solutions for the above electrodynamics equations.As pointed out before by Carroll, Fields and Jackiw \cite{4} with the introduction of a mass for the photon,geomagnetic data can place limits on the CS parameter $m_{\gamma}=(p_{\alpha}p^{\alpha})^{\frac{1}{2}}$ or $m_{\gamma}< 6X 10^{-26} GeV$. Since $p_{\alpha}$ is proportional to the torsion vector $S^{\alpha}$ the photon mass can be expressed in terms of the torsion axial vector as $m_{\gamma}=2(S_{\alpha}S^{\alpha})^{\frac{1}{2}}$. This allows us to give an estimate to the torsion vector based on the Lorentz violation parameter. Note indeed that since $p_{0} {\alpha}10^{-42} h_{0} GeV$ and $|\vec{p}| {\alpha} 10^{-42} h_{0} Gev$ and $5X 10^{-1} < h_{0} < 1 $ one obtains from expression (\ref{4}) above that torsion cosmological background is given by $ 10^{-33} eV < |S^{\mu}| < 10^{-32} eV $ which is well within the limits obtained previously. As pointed out by Obukhov \cite{11} the polarization rotational effect can be obtained by the expression
\begin{equation}
k^{\mu}{\nabla}_{\mu}f^{\nu}=0
\label{10}
\end{equation}
where $k^{\mu}$ is the wave vector while $f^{\mu}$ is the polarization vector orthogonal to the wave vector and therefore obeying the equation $k^{\mu}f_{\mu}=0$.Here ${\nabla}^{\mu}$ represents the Riemann-Cartan covariant derivative with torsion.By expanding the covariant derivative operator in terms of torsion vector shows that torsion vector will interact with the polarization and wave vectors therefore contributing to Nodland-Ralston birefrigence. A manifestly gauge invariant formulation of the coupling of the Maxwell equations and Einstein-Cartan gravity has recently been proposed by P.Majumdar and S. SenGupta \cite{1} where spacetime torsion originates from a massless Kalb-Ramond field augmented by suitable $U(1)$ CS terms.Another interesting extension of the ideas discussed here would be to extend it to the Riemann-Cartan spacetime with torsion and applies it to an specific cosmological model like the Friedmann or Bianchi IX ones.

\section*{Acknowledgement}
We are very much indebt to Professors Yuri Obukhov and I.Shapiro for carefully reading the manuscript and for helpfull discussions.  Financial support from CNPq. and UERJ is acknowledged.


\begin{thebibliography}{11}
\bibitem{1} P. Majumdar and S. SenGupta, Parity Violating Gravitational Coupling of Electromagnetic Fields,gr-qc/9906027v2. 
\bibitem{2} S. Mohanty and U. Sarkar,Constraints on Background torsion fields from K physics,(1998)ICTP preprint (IC/98/34).
\bibitem{3} A. Dobado and A.Maroto,Mod. Phys.Lett.A.
\bibitem{4} S.M.Carroll,G.B.Field and R.Jackiw,Phys. Rev.D 41,4,1231 (1990).
\bibitem{5} B. Nodland and J. P. Ralston,Phys. Rev. Lett. 78, 3043 (1997).
\bibitem{6} S.M. Carroll and G.B. Field,Is there evidence for cosmic anisotropy in the polarisation of distant radio sources? astro-ph/9704263.
\bibitem{7} D. J. Eisenstein and E.F. Bunn, astro-ph/9704263. 
\bibitem{8} J. P. Leray,astro-ph/9704285.
\bibitem{9} A.R. Prasanna and S. Mohanty, Astrophysical signals of P and T violation by gravity,astro-ph/9509025.
\bibitem{10} L.C.Garcia de Andrade,Cosmic Rotation, Birefrigence and Axions to detect Primordial Torsion Fields,hep-th/0110150.
\bibitem{11} Yu N. Obukhov in Colloquium on Cosmic Rotation,M. Scherfner (Editor), T. Chrobok and M. Shefaat (co-editors),(2000) Wissenschaft and Technik Verlag.
\end{thebibliography}
\end{document}